\documentclass[aps,floatfix,showpacs,prb]{revtex4}
\usepackage{graphicx,amsfonts,amssymb,enumerate,color}
\usepackage{natbib}

\newcommand{\be}{\begin{equation}}
\newcommand{\ee}{\end{equation}}
\newcommand{\bea}{\begin{eqnarray}}
\newcommand{\eea}{\end{eqnarray}}
\newcommand{\beb}{\begin{eqnarray*}}
\newcommand{\eeb}{\end{eqnarray*}}

\newcommand{\phrb}[3]{Phys. Rev. B{\bf #1}, #2 (#3).}
\newcommand{\phrl}[3]{Phys. Rev. Lett. {\bf #1}, #2 (#3).}
\begin{document}
\title{Shape of the magnetoroton at $\nu=1/3$ and $\nu=7/3$ in real samples}

\author{Thierry Jolicoeur}
\affiliation{Laboratoire de Physique Th\'eorique et Mod\`eles statistiques, 
CNRS, Universit\'e Paris-Sud,
Universit\'e Paris-Saclay, 91405 Orsay, France}

\date{October, 2016}
\begin{abstract}
We revisit the theory of the collective neutral excitation mode in the 
fractional quantum Hall effect at Landau level filling fractions $\nu=1/3$ and 
$\nu=7/3$. We 
include the effect of finite thickness of the two-dimensional electron gas and 
use extensive exact diagonalizations in the torus geometry. In the lowest 
Landau level the collective gapped mode a.k.a the magnetoroton always merges in
the continuum in the long-wavelength limit. In the second Landau level
the mode is well-defined only for wavevectors smaller than a critical value 
 and disappears in the continuum beyond this point. Its curvature near zero 
momentum is opposite to that of the LLL. It is well separated from the 
continuum even at zero momentum and the gap of the continuum of higher-lying 
states is twice the collective mode gap at $k=0$. The shape of the dispersion 
relation survives a perturbative treatment of Landau level mixing.
\end{abstract}
\maketitle

\section{Introduction}
Quantum liquids display very special collective 
modes that dominates their low-energy long-wavelength behavior. In the case of 
liquid Helium-4 the ground state has broken gauge symmetry associated with 
number conservation and as a consequence there is a phonon branch of 
excitations with no gap. While the gapless nature of the mode is dictated by 
general requirements, here the Goldstone theorem, its shape for finite 
wavevector is non-universal. Remarkably it has a nontrivial minimum dubbed the 
roton. This roton state has studied experimentally in great detail and 
Feynman~\cite{Feynman} has developed the so-called single-mode approximation 
(SMA) that captures in a neat way its nature.

Two-dimensional (2D) electronic systems under a strong 
magnetic field exhibit the fractional quantum Hall effect (FQHE) at low
enough temperatures. The most prominent of these states of matter happens for 
filling factor $\nu=1/3$ of the lowest Landau level~\cite{Laughlin} (LLL). It 
has been observed for conventional
semiconductor artificial devices, quantum wells and heterostructures, 
as well as in atomically 2D systems like monolayer and bilayer graphene.
The ground state of the 2D electrons for Landau level filling factor $\nu=1/3$
is adequately described by the Laughlin wavefunction. This state has no broken 
symmetry and is a prime example of topological order. The incompressibility 
that is responsible for the macroscopic phenomenology of the state also
leads to gapped collective neutral excitations. The lowest-lying density mode
can be also be described by a Landau-level adapted single-mode 
approximation~\cite{Girvin85,Girvin86} (SMA)
and it features also a minimum energy as a function of wavevector. This
minimum is called the magnetoroton. In addition to the SMA its mere existence 
has been confirmed by exact diagonalization of small systems and there are trial
wavefunctions constructed with composite-fermion states. 
In the composite fermion approach the magnetoroton is the lowest-energy 
particle-hole excitation between effective composite fermion Landau 
levels~\cite{JainBook,Toke05}. Other wavefunctions have also been 
proposed~\cite{BY2012,BY2013,BY2014}. Other FQHE states have more complex
neutral excitations~\cite{Balram2016,Golkar}.
Inelastic light 
scattering has been used to probe the collective mode and is partly explained
by existing theories~\cite{PinczukReview,Pinczuk93,Pinczuk98,Platzman94,He96}. 
One intriguing 
suggestion is the existence of a two-roton bound 
state~\cite{Park2000,Ghosh,He96} as a possible lowest-lying state for small 
wavevectors. This question has proven difficult to answer mainly because of 
limitations of exact diagonalization to very small systems.

In this paper we  study the Laughlin state at $\nu=7/3$ and obtain the neutral 
excitation spectrum by using large-scale exact 
diagonalizations on the torus geometry with spin-polarized electrons.
The transport phenomenology of this fraction has been known 
for some time to be analogous to that of $\nu=1/3$ as theoretically expected 
by uplifting
the Laughlin wavefunction in the second Landau level~\cite{MacDonald84} (LL). 
Here 
we find 
that there are very strong finite size effects, obscuring the spectrum structure 
up to $N_e=10$
electrons. However non-zero thickness of the 2D electron gas (2DEG) 
resuscitates 
 the familiar magnetoroton provided one reaches large enough systems with 
$N_e\geq 11$ and use the torus geometry. For small width $w/\ell\lesssim 1$ 
(where $\ell$ is the magnetic length $\sqrt{\hbar/eB}$)
 the system may be  compressible but for $w/\ell = 2-3$ we observe the clear 
signature of the MR mode.
However it has now
 a different structure w.r.t its LLL counterpart~: it is well-defined only for 
wavevectors $k\ell\lesssim 1.8$
and enters the continuum beyond this value. For $k\approx 0$ there are two 
 gaps leading again to a well-defined mode in the long-wavelength limit. 
Contrary to the LLL case the curvature of the dispersion relation is upwards 
close to $k=0$
and there is a secondary maximum in addition to a roton minimum. There is no 
clear limiting behavior when $k\rightarrow\infty$ which is in line with the 
fact that in this limit the magnetoroton is expected to become a 
quasihole-quasielectron pair and their size is probably very 
large~\cite{Balram2013,Johri2014}. It may very well be that the collective mode 
at small
wavevector is not continuously connected to the quasiparticle-quasihole mode 
expected at larger wavevector. The shape of the dispersion relation of MR mode
is essentially unaffected by Landau level mixing effects, at least in a 
perturbative treatment.

In section II we present several models to take into account the finite 
thickness of the 2DEG in the Coulomb interaction potential. In section III we 
use exact diagonalization on the torus geometry to study the LLL magnetoroton.
Section IV is devoted to the shape of the MR at $\nu=7/3$, it contains an 
extensive discussion of finite-size effects with the torus and sphere geometry.
Landau-level mixing is treated perturbatively in section V. Finally section VI 
contains a discussion of experimental results on the MR shape and our 
conclusions.

\section{Model interactions for finite-width effects}

The wavefunctions of electrons in 2DEG samples have a finite extent in the $z$ 
direction perpendicular to the plane of the electron gas. However the 2D nature 
of the motion means that excitations in this $z$ direction are energetically 
forbidden~: the $z$ motion is frozen in its ground state. One can thus make the 
approximation that electronic wavefunctions factorizes. If we call $\phi_0$ the 
ground state for the $z$ coordinate then the
 electron interactions can be written as~:
\be
V(\mathbf{x}-\mathbf{y})=\frac{e^2}{\epsilon}
\int dz_1 dz_2 \frac{|\phi(z_1)|^2 
|\phi(z_2)|^2}{\sqrt{(\mathbf{x}-\mathbf{y})^2+(z_1-z_2)^2}}.
\label{zpot}
\ee
This modified form of the potential can then be written in second quantized 
form projected onto the LLL or the second Landau level.
The Fourier transform of the interaction potential has now the following form~:
\be
\tilde{V}(q)=\frac{2\pi e^2}{\epsilon q}
\int dz_1 dz_2\,\, |\phi(z_1)|^2 
|\phi(z_2)|^2 \,\,{\mathrm{e}}^{-q|z_1-z_2|},
\label{fourier}
\ee
where we see that the bare Coulomb potential $2\pi e^2/q$ is multiplicatively 
changed by a form factor depending upon the shape of the $z$ wavefunction.
2DEGs comes mostly in two varieties~: heterostructures and quantum wells. In 
the first 
case the potential felt by electrons in the $z$ direction perpendicular to 
the 2D plane is approximately 
triangular and a trial wavefunction has been proposed by Fang and Howard~:
\be
\phi_{FH}(z)=\frac{2}{(2b)^{3/2}} ze^{-bz/2},
\label{FHw}
\ee
where the parameter $b$ can be determined variationally as a function of the 
junction parameters including the electronic density~\cite{Ortalano97}. In 
quantum wells (QW)
the corresponding wavefunction is the usual eigenstate for a square well
with some finite width $d$~:
\be
\phi_{SQ}(z)=\sqrt{\frac{2}{d}} \cos (\frac{\pi z}{d}).
\label{SQw}
\ee
It has also been suggested that one can 
use a simple ad hoc modification of the potential at short distances. Zhang and 
Das Sarma have proposed the substitution~:
\be
\frac{1}{r}\rightarrow \frac{1}{\sqrt{r^2+a^2}},
\ee
where the cut-off $a$ has the dimension of a length and captures the 
finite-width effect.
From a practical point of view it has been noted that a simple Gaussian 
wavefunction reproduces correctly the LL-projected Hamiltonian~\cite{Morf2002}~:
\be
\phi_{GS}(z) = \frac{1}{(w\sqrt{2\pi})^{1/2}}{\mathrm{e}}^{-z^2/4w^2}.
\label{GSw}
\ee
This is the  model we use in this paper~: we quote the width $w$ without
extra subscript when it corresponds to the Gaussian model while we add the name 
of the model wavefunction as a subscript otherwise.
The effect of the finite width has been studied notably by exact 
diagonalization of small systems~\cite{Mike08,Mike08L,Morf2002}.
It is known that wide quantum wells have a tendency to stabilize quantum Hall 
states
in the second Landau level mainly based on overlap calculations with model 
states~\cite{Mike08,Mike08L,Papic2009}.
If the width of the well becomes very large then the electronic density may 
prefer to
form two layers of charge giving rise to an effective two-component system 
whose physics is now quite distinct~\cite{Shayegan98}. In this work we consider 
only the single-component case which is realized for not too thick samples.
 It is known that there 
are samples with density small enough to stay in the one-component regime
but with values $w/\ell$ up to 5 to 6 for $\nu=1/3$. Previous studies have 
shown that the activation gap decreases with $w$ but the FQHE regime 
survives~\cite{Shayegan90,Suen94,Manoharan96,He90,Halonen93}.
For example references (\onlinecite{Manoharan96,Shayegan98}) have studied a 
system
with density $n=6.4\times 10^{-10}$ $cm^{2}$ and a width $w_{SQ}=75$nm which 
leads to
a fractional quantum Hall state at $\nu=7/3$ with $w_{SQ}/\ell=3.2$ while 
retaining  
single-component physics.

\section{the LLL magnetoroton}

\begin{figure}[t]
 \includegraphics[width=0.5\columnwidth]{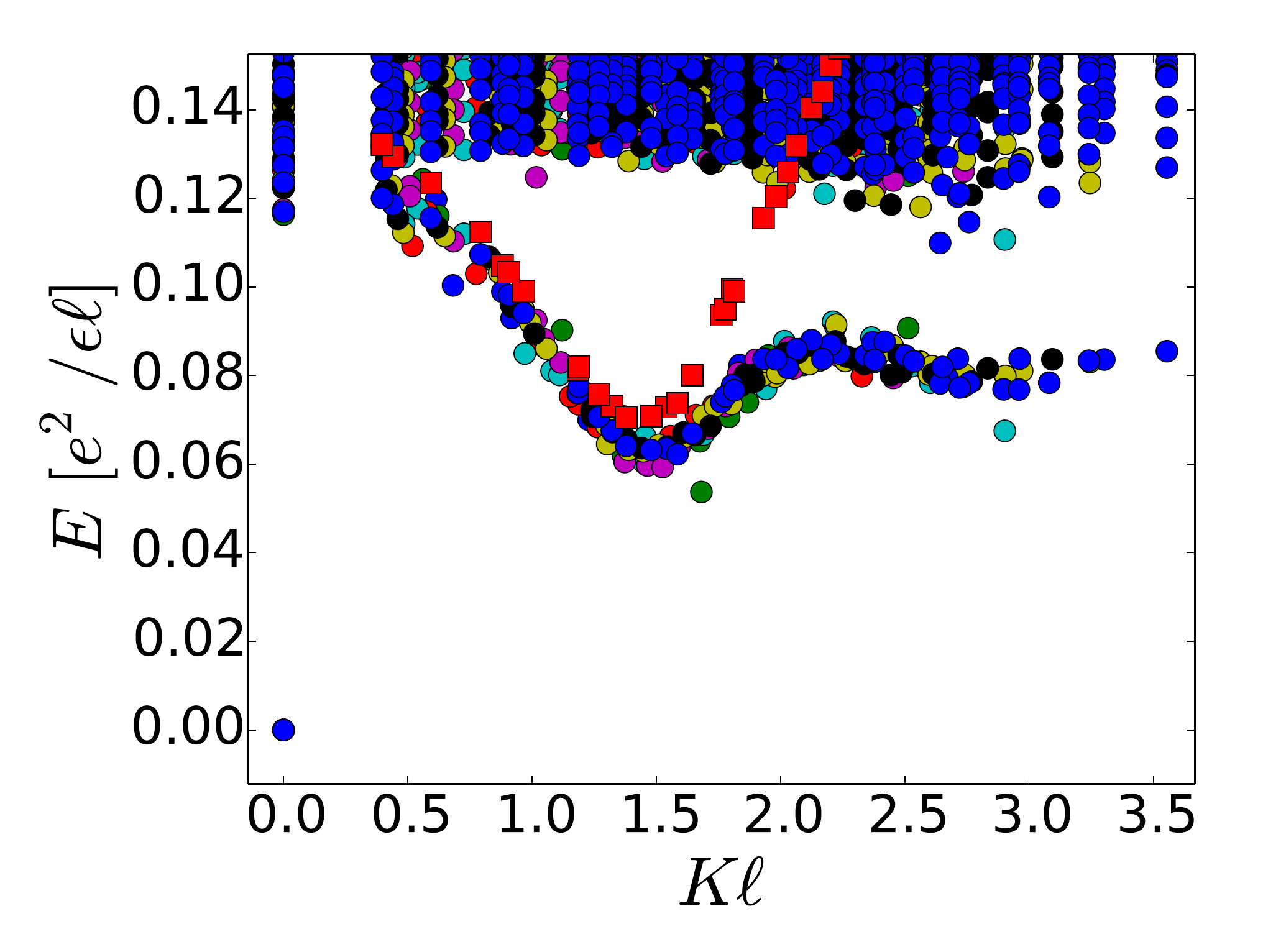}
 \caption{Spectra for systems of $N_e=5$ up to $N_e=12$ electrons 
at filling $\nu=1/3$ in a 
rectangular cell with aspect ratio $L_x/L_y=0.9$. No finite width is assumed.
The magnetoroton branch is clearly defined for all momenta. There is a clear 
minimum at wavevector $K\ell\approx 1.5$. Finite-size 
effects are negligible. The biggest system corresponds to blue points.
The red square points are the SMA values.}
 \label{LLL}
\end{figure}

\begin{figure}
\begin{center}
  \includegraphics[width=0.4\columnwidth]{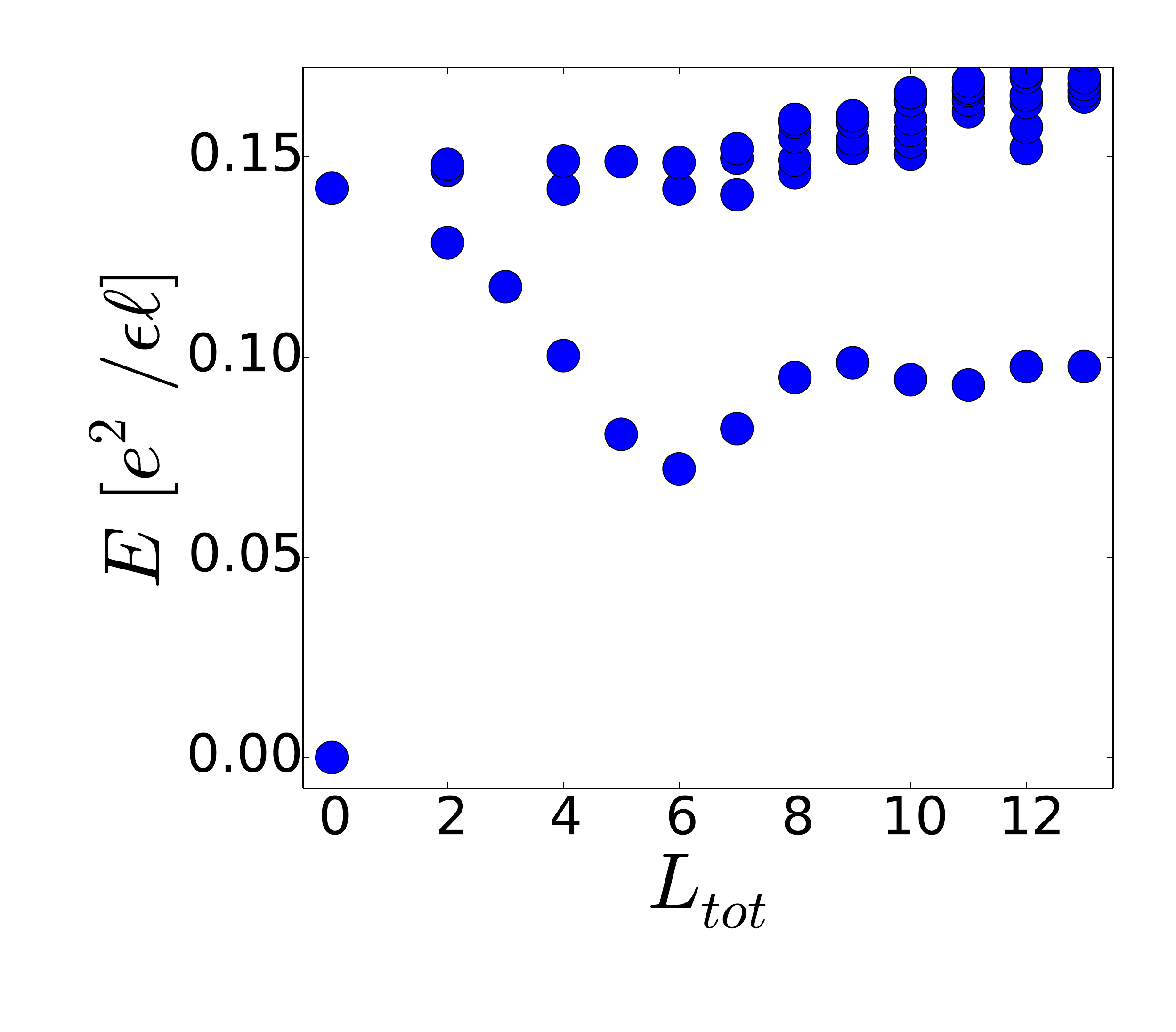}
  \caption{Sphere spectrum for $N_e=13$ electrons in the LLL with pure Coulomb 
interaction. The magnetoroton branch is well defined and extend up to 
$L=N_e=13$.
The shape of the magnetoroton is similar to the torus case in Fig.(1).}
  \label{S1}
\end{center}
\end{figure}

To study the FQHE with the modified 
potential Eq.(\ref{zpot}) we use the torus geometry~\cite{Haldane87}~:
magnetic 
translation symmetries allow to classify eigenstates by a two-dimensional 
conserved quasimomentum $\mathbf{K}$
living in a discrete Brillouin zone with only $N^2$ points where $N$ is the GCD 
of the number of electrons $N_e$ and the number of flux quanta $N_{\phi}$
though the system. We fix the filling factor $N_e/N_{\phi}=1/3$.
On a rectangular cell with periodic boundary conditions we 
have~:
\be
\mathbf{K}^2=(2\pi s/L_x)^2+(2\pi t/L_y)^2,
\label{mom}
\ee
where $s,t$ are integers running from 0 to $N$. The many-body eigenstates can 
then be plotted against the dimensionless momentum $k\ell\equiv 
|\mathbf{K}|\ell$. We use an aspect ratio $L_x/L_y=0.9$ in this work since 
the physics is only weakly dependent of this value even in the second LL.
For zero width we recover the well-known shape of the magnetoroton 
mode~\cite{Girvin85,Girvin86}.  For small wavevectors its energy rises and 
disappears into the continuum. He and Platzman have argued~\cite{He96}
that there is a crossing of levels close to $k=0$ and that a state with a 
two-roton character becomes lower in energy. Present data does not shed any 
light on this issue.  The magnetoroton appears clearly in Fig.(\ref{LLL}). The 
ground 
state is at $\mathbf{K}=0$ and is isolated by a gap from all excitations~:
this gap is larger than the typical level spacing to the finite number of 
particles.
The magnetoroton branch is well-defined and is rather insensitive to 
finite-size effects. 
When including finite-width effects the overall shape does 
not 
change, notably the $\mathbf{K}=0$ behavior remains but the energy scale is 
lowered. This is the case for realistic wavefunctions included in the modified 
potential Eq.(\ref{zpot}), be it Gaussian, square well or Fang-Howard,
provided that the charge 
distribution has the same variance, spectra are very similar and differ only
in quantitative details.
To get a trial wavefunction to 
describe the 
MR, Girvin MacDonald and Platzman have proposed to adapt an idea due to Feynman.
One creates a density excitation by acting upon the ground state with a density 
operator of definite momentum $\hat\rho_{\mathbf{K}}$ and one projects the 
resulting state into the LLL~:
\be
|\Psi_{SMA}(\mathbf{K})\rangle\equiv 
\mathcal{P}_{LLL}\,\, \hat\rho_{\mathbf{K}}|\Psi_0\rangle .
\label{SMAdef}
\ee
These trial states give a successful estimate of the energy of the MR in the 
LLL case~: see Fig.(\ref{LLL}) where we have plotted the SMA energy estimates 
in red 
squares. The small 
wavevector behavior and the MR minimum are correctly reproduced. 
In the CF theory it is possible to build excitonic states in which a composite 
fermion is raised from the lowest CF Landau level into the first excited CF 
Landau level. This also gives a satisfactory description of the MR mode. We 
note 
that sphere calculations also reveal the presence of the MR mode. In 
Fig.(\ref{S1}) we plot the eigenenergies of $N_e=13$ electrons at $\nu=1/3$
as a function of the total angular momentum $L_{tot}$. The MR branch is also 
prominent
and extend up to $L_{tot}=N_e$ as predicted by CF theory on the sphere.

\begin{figure}
\begin{center}
  \includegraphics[width=0.4\columnwidth]{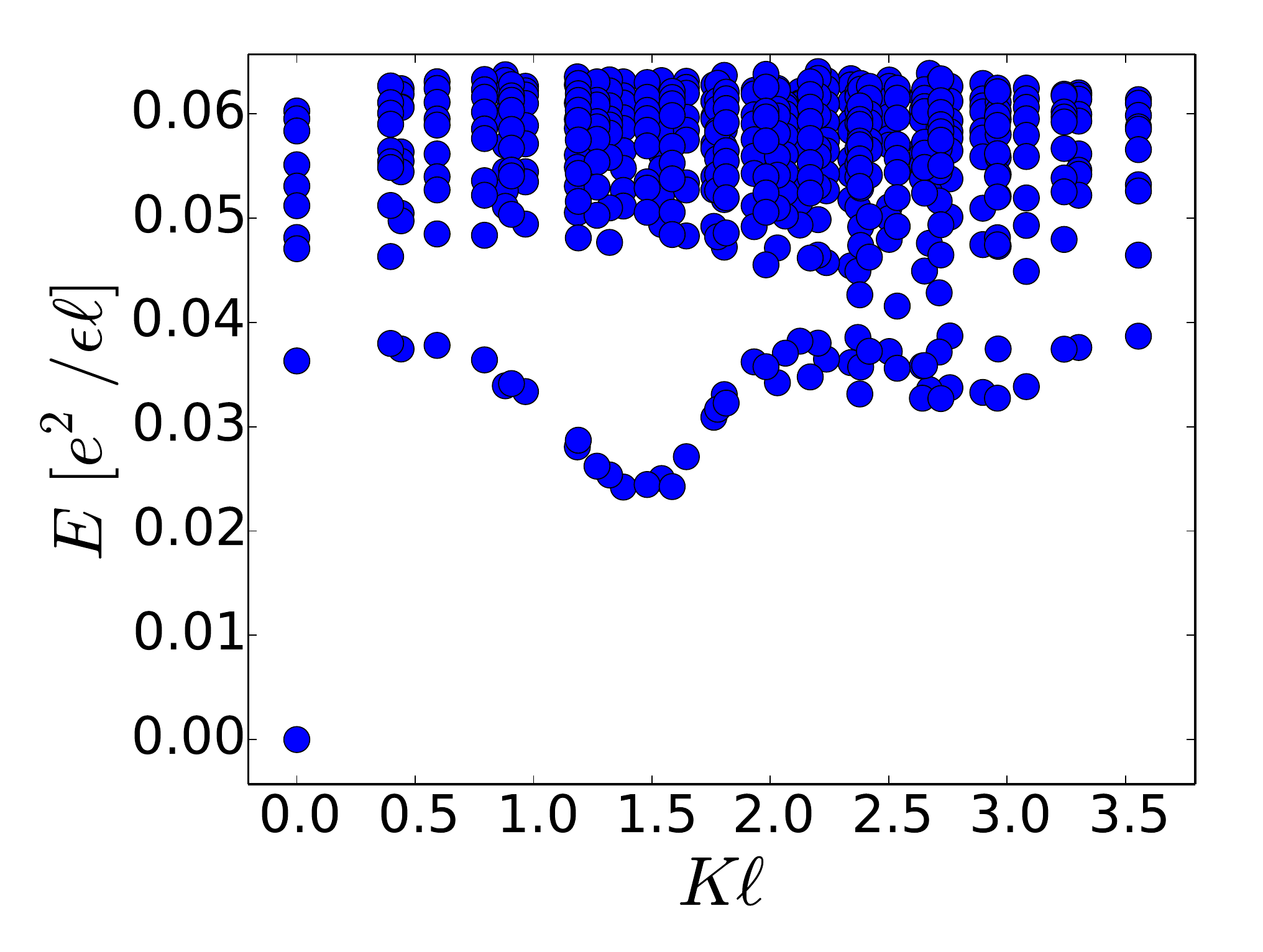}
  \caption{Torus spectrum for $N_e=12$ electrons in the LLL. The aspect ratio 
is 
0.9. The finite-width is now captured by the Zhang-Das Sarma modification
of the Coulomb potential with parameter $a=4\ell$. This has a big impact on the 
MR shape.
Notably it does not enter the continuum in the long-wavelength limit.}
  \label{S4}
\end{center}
\end{figure}

If we consider a realistic potential with finite-width we find that the shape 
of the MR mode is unchanged for the cases of Fang-Howard, square well and 
Gaussian 
models. However
this is not the case with Zhang-Das Sarma proposal~\cite{ZDS} 
which has a qualitative impact in the LLL. For example with $a=4\ell$
the LLL MR is split off from the continuum at $K=0$~:~see Fig.(\ref{S4}).
This example show that some of the MR features are non-universal~: there is 
no reason why the collective mode branch should merge in the continuum at small 
wavevectors. 

\section{Second Landau level}
\begin{figure}[h]
\begin{center}
  \includegraphics[width=0.4\columnwidth]{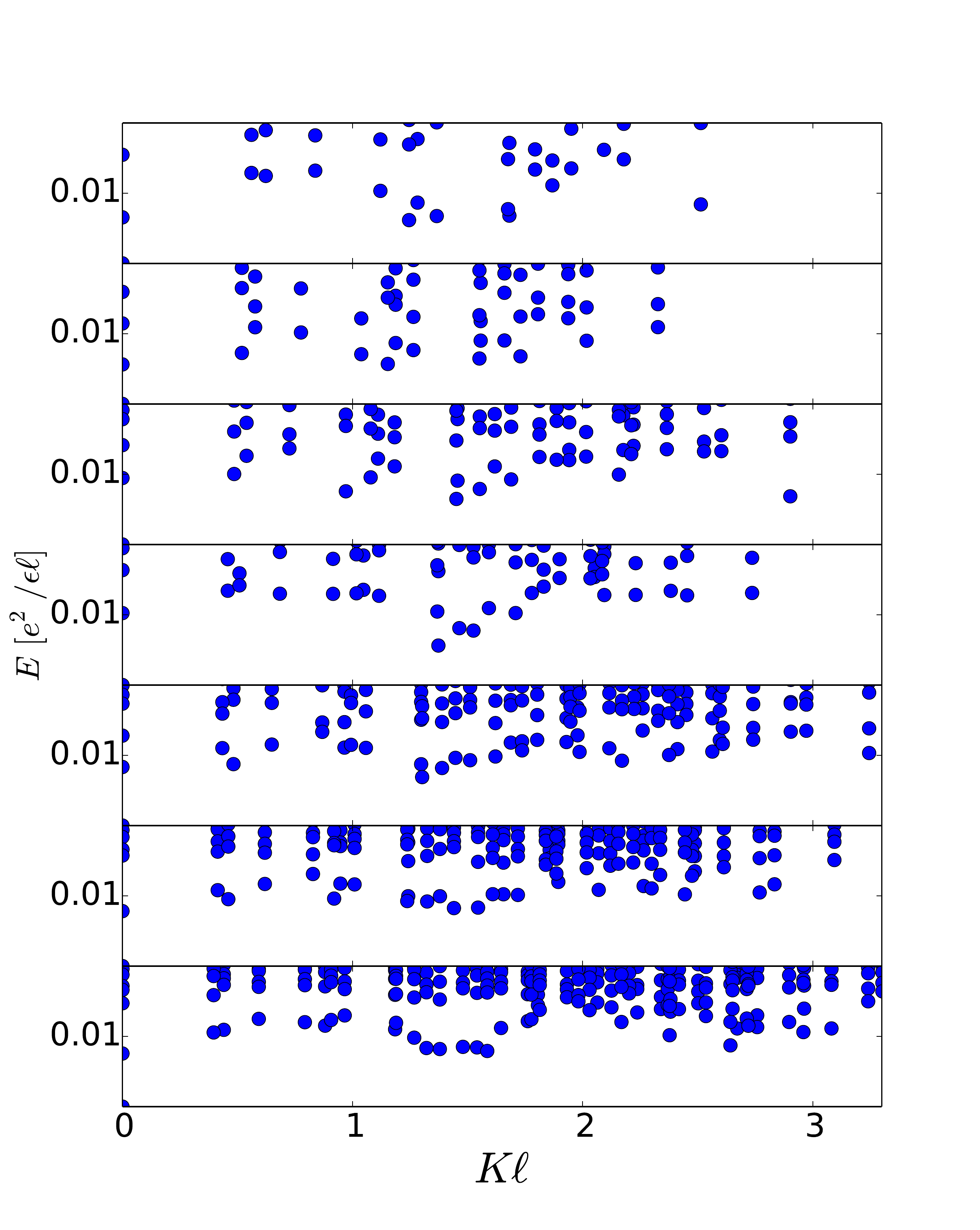}
  \caption{Evolution of the torus spectrum when increasing the number of 
particles. From top to bottom $N_e=6$ to $12$. The cell is rectangular with 
aspect ratio 0.9 and we use a finite-width potential $w/\ell=3$. The collective 
mode branch is discernible for $N_e=10$ and fully developed only for $N_e=12$.}
  \label{T1}
\end{center}
\end{figure}

We now turn to the study of $\nu=7/3$ 
state. 
The first approach to the FQHE physics in the second Landau level is to use the 
appropriate
orbital wavefunctions and fully neglect Landau level mixing. This is an 
approximation which is certainly less good than in the LLL but has the merit of 
mapping 
the problem on the very same Hamiltonian as in the LLL with renormalized matrix 
elements. This is the point of view we adopt in this section with the added 
finite-width 
effects. We will consider perturbative treatment of LL mixing in section 
\ref{LLmix}.
For small systems it is known\cite{Haldane87} 
that the system is probably gapless in the zero-width case at filling factor 
$\nu=7/3$. While the study of 
larger system sizes up to $N_{e}=12$  does not allow to strengthen this 
conclusion we find 
that nonzero width gives rise to a well-defined collective mode well separated 
from the continuum.
We observe that
the MR branch appears beyond approximately 10 electrons as can be seen in 
Fig.(\ref{T1}).

\begin{figure}[t]
 \includegraphics[width=0.4\columnwidth]{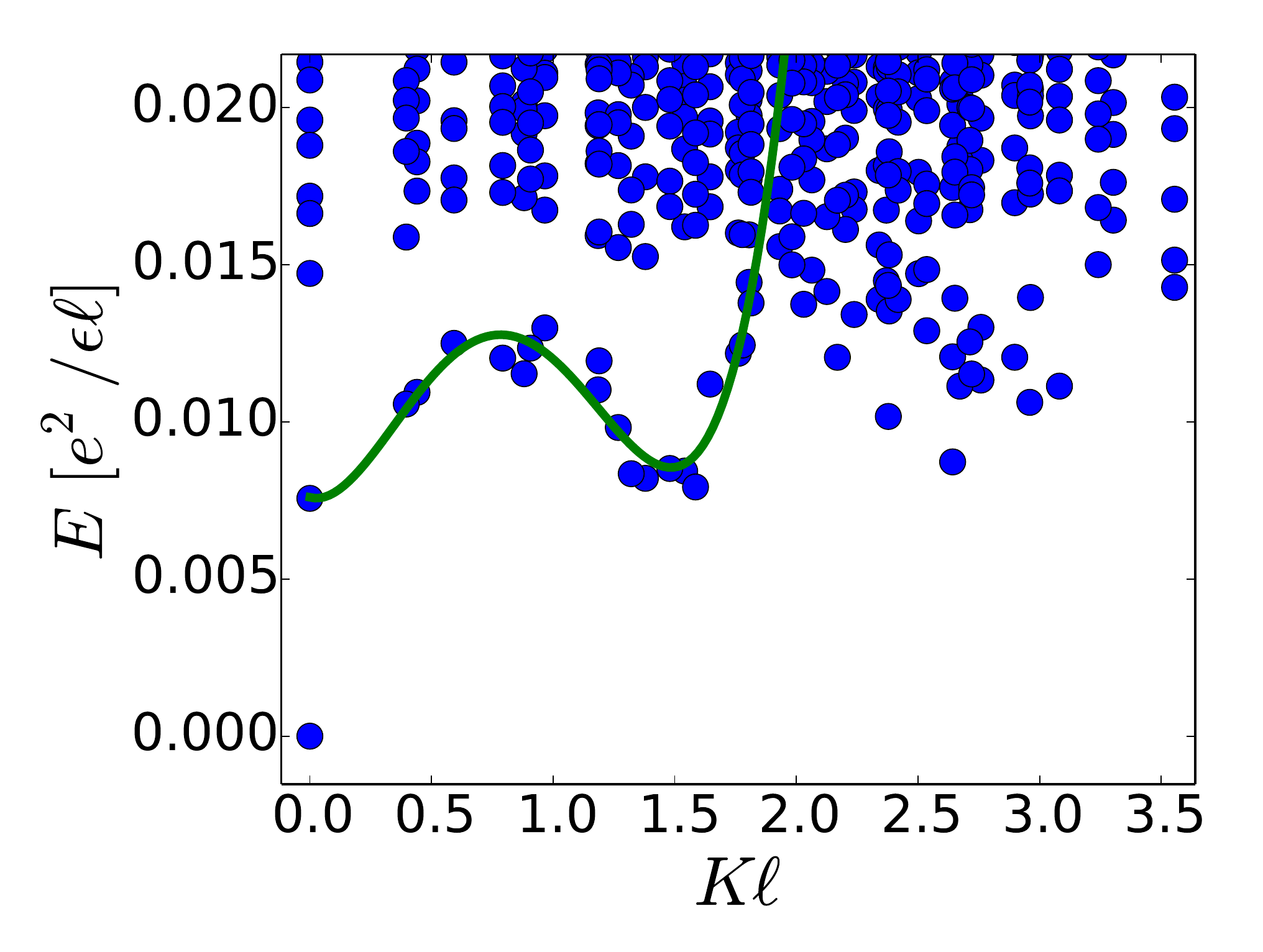}
 \caption{Spectrum of $N_e=12$ electrons on a rectangular cell with aspect 
ratio $L_x/L_y=0.9$ for $\nu=7/3$. The width is taken to be Gaussian $w/\ell=3$.
The magnetoroton branch is clearly defined only for momenta $\lesssim 
1.8\ell^{-1}$. The solid green line is a fit by a quartic polynomial. The 
dispersion relation has a maximum and a minimum.}
 \label{torus}
\end{figure}

\begin{figure}[t]
\includegraphics[width=0.4\columnwidth]{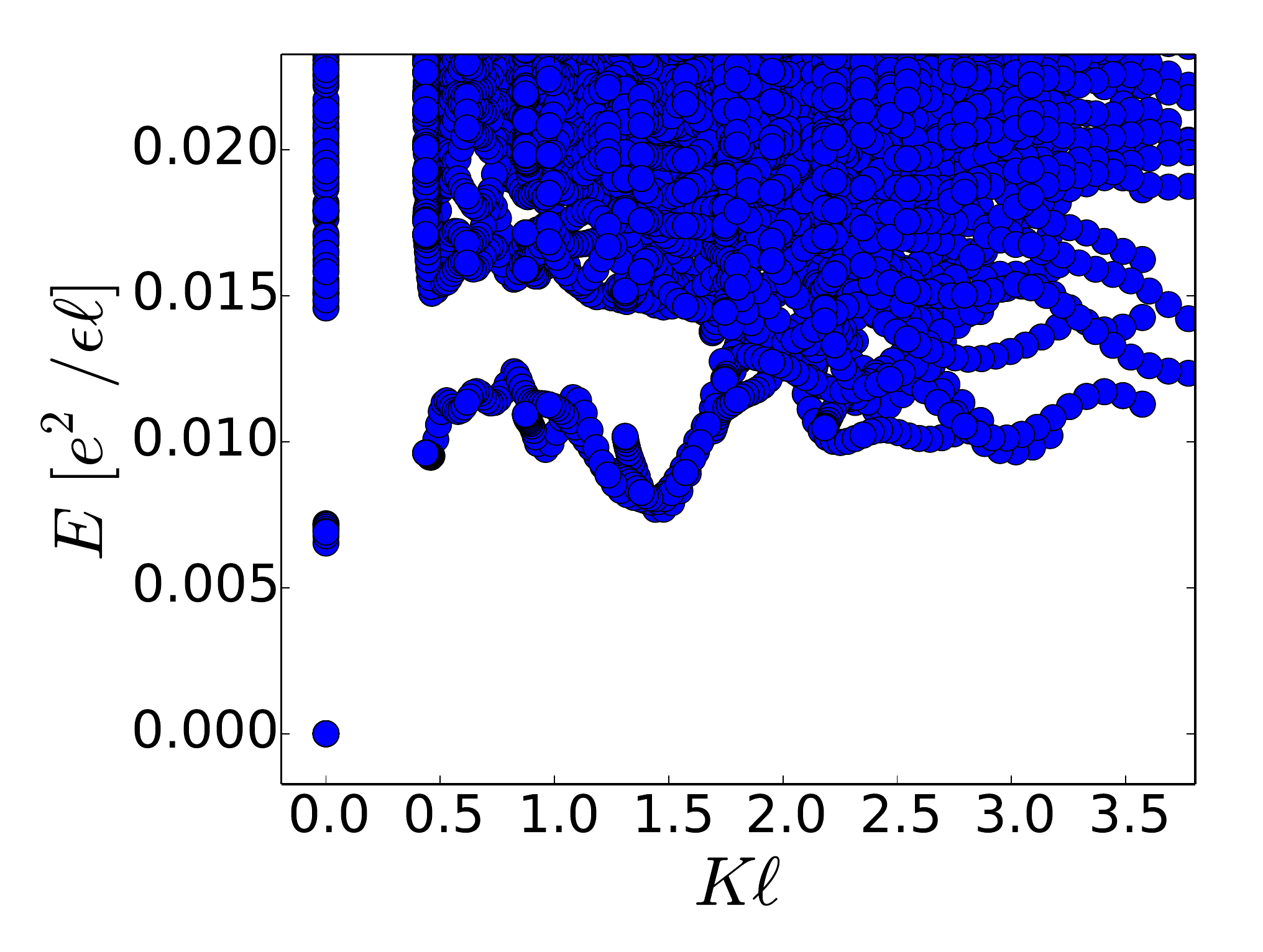}
 \caption{Excitation spectrum for $N_e=11$ electrons in an oblique cell with 
equal sides and an angle that interpolates between square and hexagonal cell.
Spectra for all shapes are plotted in this figure~: this is a way to obtain 
more points in the dispersion relation of the collective mode.
The potential assumes a Gaussian width $w/\ell=3$.}
\label{obliqcell}
\end{figure}

\begin{figure}[t]
 \includegraphics[width=0.4\columnwidth]{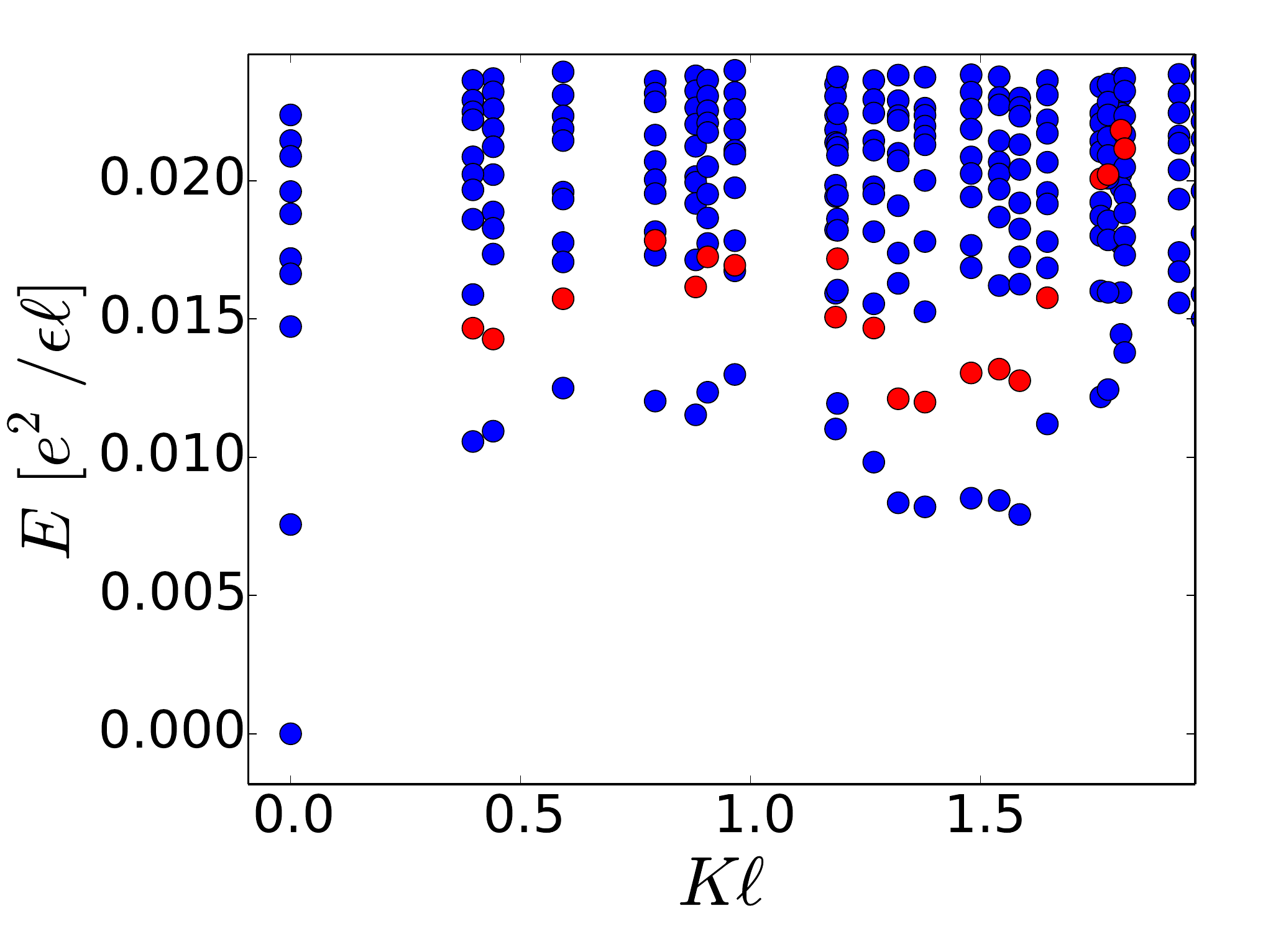}
 \caption{Spectrum of $N_e=12$ electrons on a rectangular cell with aspect 
ratio $L_x/L_y=0.9$ for $\nu=7/3$ and Gaussian width $w/\ell=3$. Blue points 
comes
from exact diagonalization while red points are results of the SMA applied to 
the exact ground state. The shape of the collective mode is correctly 
reproduced but the energies are much higher.}
 \label{SMA2LL}
\end{figure}

\begin{figure}
  \includegraphics[width=7cm]{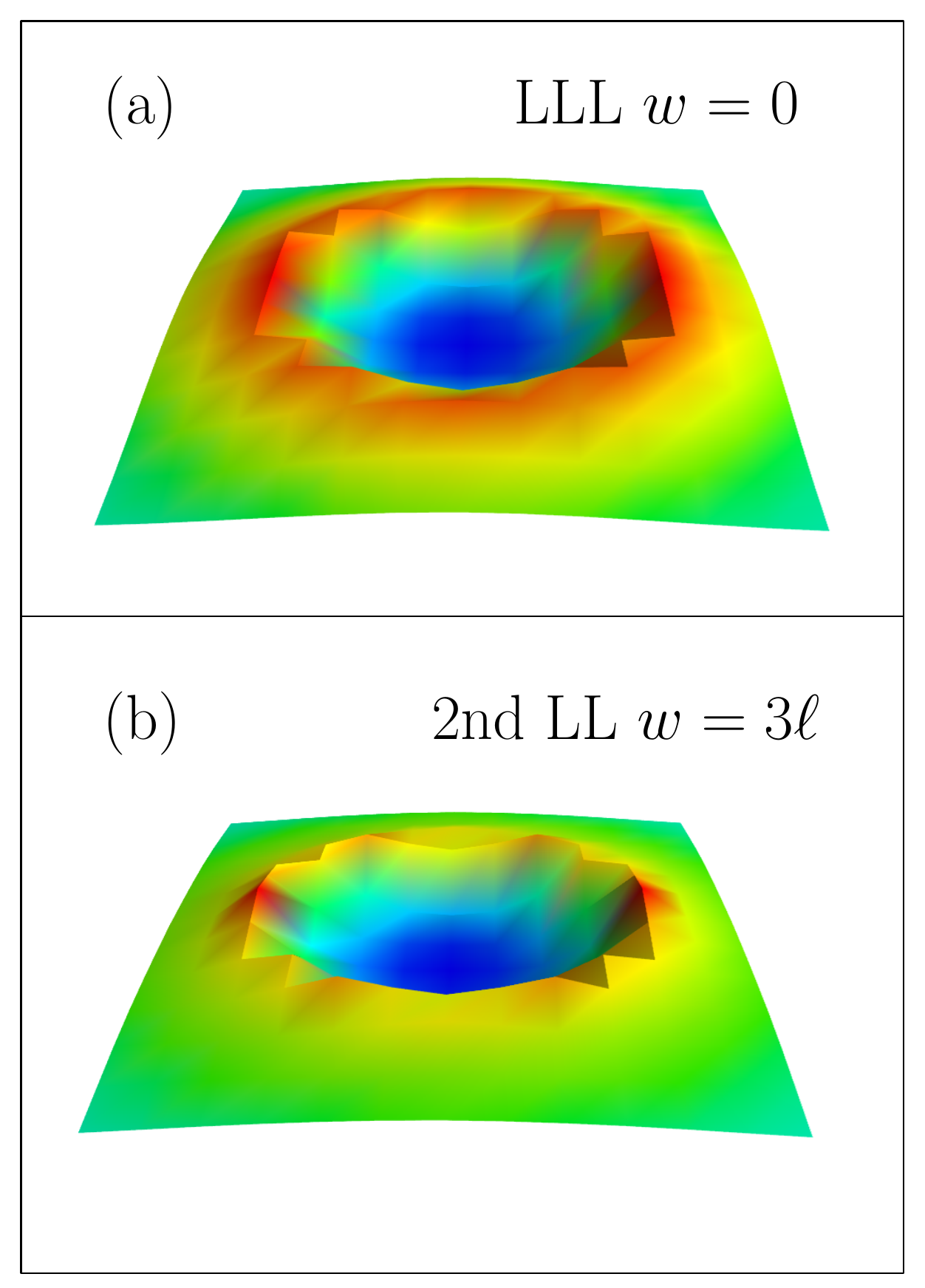}
  \caption{The guiding center structure factor $S_0(\mathbf{q})$ for $N_e=12$ 
electrons in a rectangular cell with aspect ratio $L_x/L_y=0.9$ for the 
$\nu=1/3$ state (a) in the LLL and zero width. In the second LL (b) with a 
Gaussian width $w/\ell=3$ the isotropy is almost as good as in the LLL.}
  \label{gcsf}
\end{figure}

We plot a typical spectrum in Fig.(\ref{torus}) for 
$N_e=12$ electrons
in a rectangular cell with a Gaussian wavefunction width $w/\ell =3$. The MR 
mode
now is definitely separated from the continuum at $\mathbf{K}=0$ and
has one maximum and one minimum before seemingly entering the continuum for
$k\ell\approx 1.8$. There is an obvious limitation in torus 
calculations which is the dramatic
discretization 
of momenta in Eq.(\ref{mom}). One way to overcome this is to perform 
calculations in an oblique cell with varying angle, allowing overlaps in 
momenta definition. This can be seen in Fig.(\ref{obliqcell})
where we have used a set of unit cells interpolating between a square and an 
hexagonal cell. One can see more clearly the dispersion relation of the MR mode.
The features that we observe in the single rectangular cell of Fig.(\ref{torus})
stand out clearly. 
Concerning the spectrum at zero wavevector, our data show that the 
\textit{second} 
excited state has an energy gap which is very close to twice the first energy 
gap. So the continuum of states is likely to be a two-particle continuum made
of the MR excitations.
By using the ground 
state wavefunction obtained by exact diagonalization for $\nu=7/3$ one can 
obtain the SMA states in the second LL. The results
for $N_e=12$ electrons are plotted in Fig.(\ref{SMA2LL}) with red symbols.
The overall shape of the collective mode is well reproduced by the SMA while 
now the energies are too high by a factor of $30-50\%$. We note that
the SMA works only if we use the ground state from exact diagonalization in 
Eq.(\ref{SMAdef})~: 
indeed using the Laughlin wavefunction in the 2nd Landau level is much less 
satisfactory~\cite{AHM86}.
An important quantity that can be derived easily from the SMA wavefunction is 
the 
LL-projected static structure factor which can be defined through the guiding 
center coordinates~$\mathbf{R}_i$~:
\be
S_0\left[\mathbf{q}\right]=\sum_{i<j}
\langle \exp{i\mathbf{q}(\mathbf{R}_i-\mathbf{R}_j ) } \rangle
\ee 
It can be used to reveal the fluid or crystalline character of the system. When 
evaluated in the $\nu=1/3$ state it is almost perfectly isotropic~: see the 
three-dimensional plot Fig.(\ref{gcsf}a). For the largest system studied here
this quantity is also isotropic for $\nu=7/3$ with no evidence of incipient 
charge-density wave order: see Fig.(\ref{gcsf}b).

The composite fermion 
wavefunctions are also able to reproduce the MR dispersion accurately in the 
LLL but only in the sphere or disk geometry. In Scarola et 
al.~\cite{Scarola00-I,Scarola00-II} there is a calculation 
of the MR in 
the 2nd LL which has a shape similar to what we observe but these results are 
obtained in the case of zero-width. It would be interesting to compare CF 
calculations of the MR mode including finite-width effects.

\begin{figure}
\begin{center}
  \includegraphics[width=0.4\columnwidth]{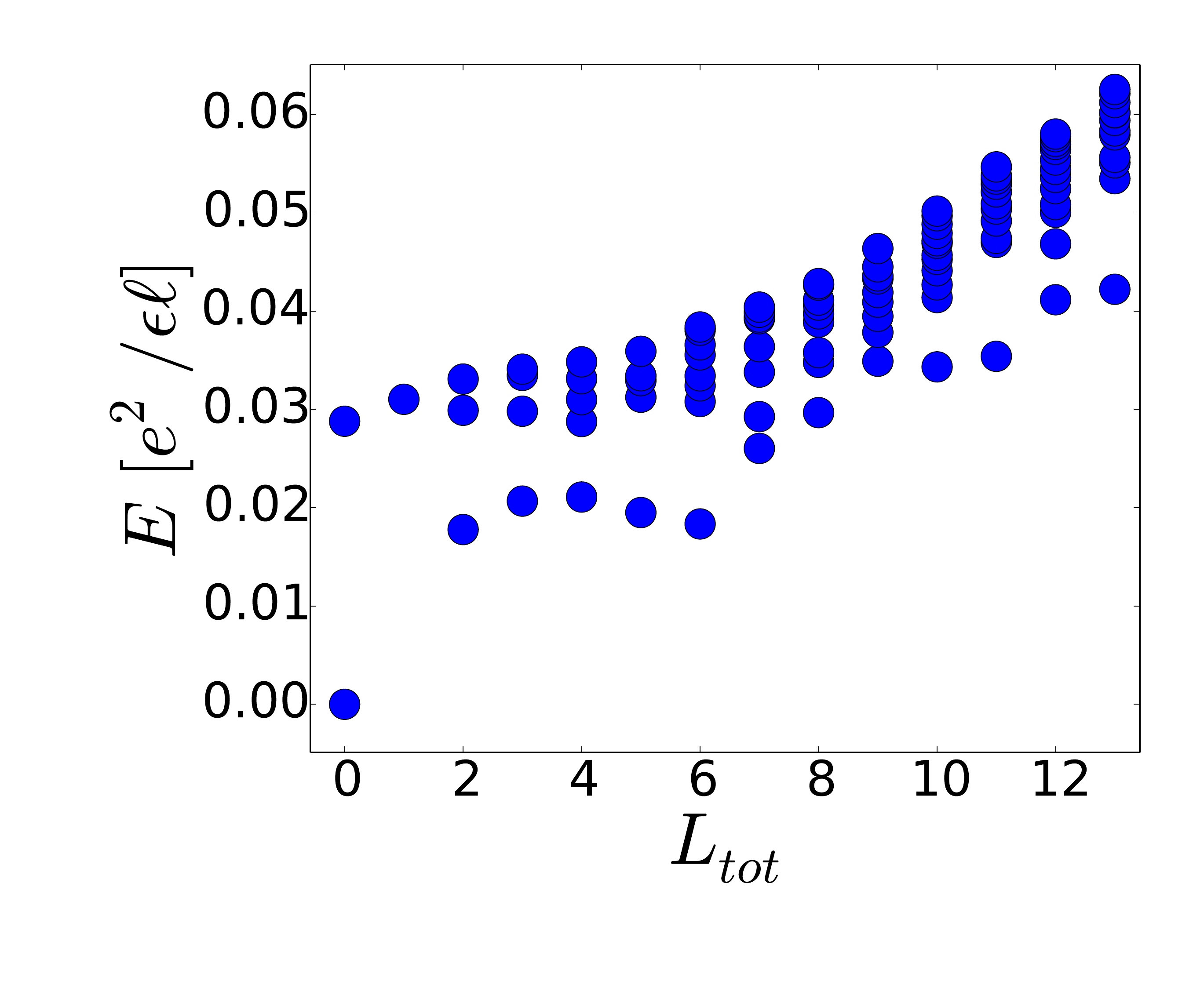}
  \caption{Sphere spectrum for $N_e=13$ electrons in the 2nd LL using planar 
pseudopotentials with finite width modeled after a square well of width 
$w/\ell=3$. The MR mode has a downward curvature for small enough momenta
as observed in the torus geometry
There are no states at low energy for $L=1$ as in the LLL case and at variance 
with the LLL it is not clear that the branch extend to a nontrivial value at 
$L=0$.}
  \label{S2}
\end{center}
\end{figure}

Not all these MR features are seen in the spherical 
geometry~\cite{Wojs2001,Wojs2009,Simion2008} for the same sizes which means that
the torus finite-size effects are different from those observed in the 
spherical geometry.
In the sphere case energy levels are classified by the total angular momentum 
and the MR branch is obvious in the LLL, extending up to $L=N_e$ as explained 
by the exciton picture of composite fermions. The LLL finite size effects are 
very 
small. Indeed the shape of the MR branch is obvious even for small systems.
When going to larger systems it is possible to observe oscillations at large 
momentum~: see Fig.(\ref{S1}).
On the sphere geometry there is no intrinsic way to treat the $z$-extent of 
the electronic wavefunction so one has to use the interaction matrix 
elements computed in the 
planar case. This adds an uncontrolled bias that will disappear only in the 
thermodynamic limit. A sample calculation for $N_e=13$ in the second LL is 
presented in 
Fig.(\ref{S2}).
Some features are in agreement with the torus results. Notably the curvature of 
the MR mode is downwards at small momentum. The mode is very close or merge
with the higher-lying continuum for $L\approx 7-8$. This corresponds to the 
characteristic wavevector $k\ell\approx L/R\approx 1.7$ where $R$ is the sphere 
radius in agreement with the torus result.
One may interpret the states
at higher L values as a quasiparticle-quasihole branch since it terminates also 
at $L=N_e$ as predicted by the exciton CF picture. The most serious discrepancy 
is the lack of low-lying states at $L=1$ and accordingly there is no state at 
$L=0$ candidate to come from the smooth continuation of the MR branch at small 
angular momentum. The low angular momentum states are those with the largest
spatial wavelength and it is certainly those states that are most sensitive to 
the curvature of the sphere.
This means that
one needs much larger spheres than toruses to capture the $\nu=7/3$ physics. 
Note that torus geometry for zero 
width in the LLL case $\nu=1/3$ gives results consistent with the sphere 
results~\cite{Morf2002}:
the large wavevector limit of the MR branch is approximately $0.09 
e^2/\epsilon\ell$ in Fig.(\ref{LLL}).

The finite-size dependences of the gaps one can define in 
the spectra above are very irregular. This is the case in the LLL but to a 
lesser degree. So it is difficult to give an estimate of gaps even for 
finite width. We just quote that for bigger sizes the $K=0$ gaps seem to 
stabilize
and are $\approx 0.017e^2/\epsilon\ell$ for $w=2\ell$ and
$\approx 0.0075e^2/\epsilon\ell$ for $w=3\ell$. As is the case for the LLL, the 
gaps are smaller with wider wells. For smaller widths $w/\ell\lesssim 1$
the MR branch is not well-defined and it is not possible to give a estimate of 
the $K=0$ for MR gaps. The gaps we estimate are defined through the dispersion 
relation of the neutral collective mode. As such they are not directly related 
to the 
quasiparticle-quasihole gap that governs the activated law of the longitudinal 
resistance.

\section{Landau level mixing}
\label{LLmix}

\begin{figure}
   \centering
 \includegraphics[width=8cm]{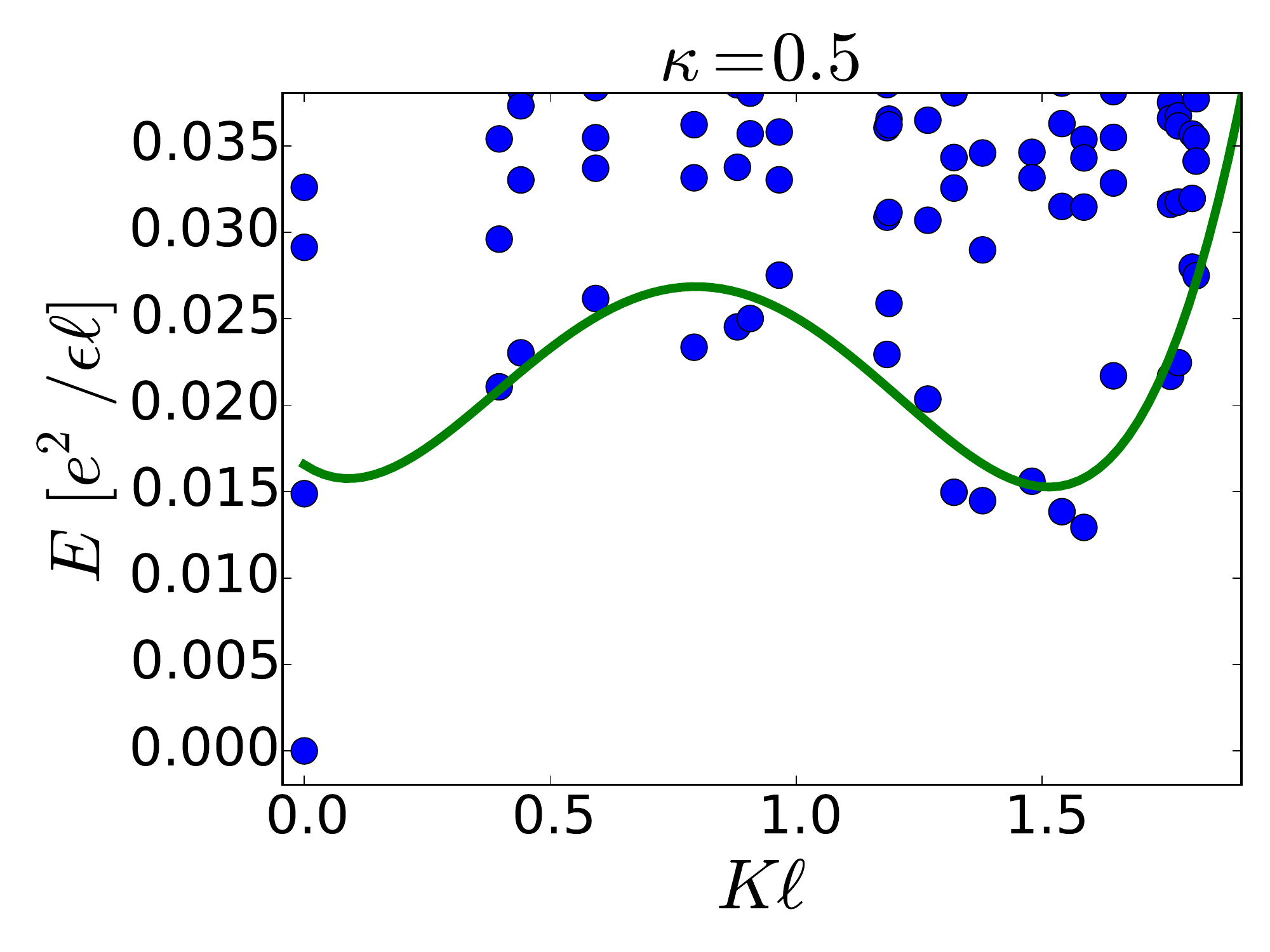}
  \caption{Torus spectrum for $N_e=12$ electrons in the 2nd LL. Aspect ratio is 
0.9. Finite-width effects are modeled by a square well $w_{SQ}=4\ell$.
The LL mixing is treated as a perturbation with $\kappa=0.5$. Close-up on the 
MR mode~:
its shape is essentially unchanged with respect to the zero LL-mixing case.}
  \label{LL05}
\end{figure}

\begin{figure}
  \centering
 \includegraphics[width=8cm]{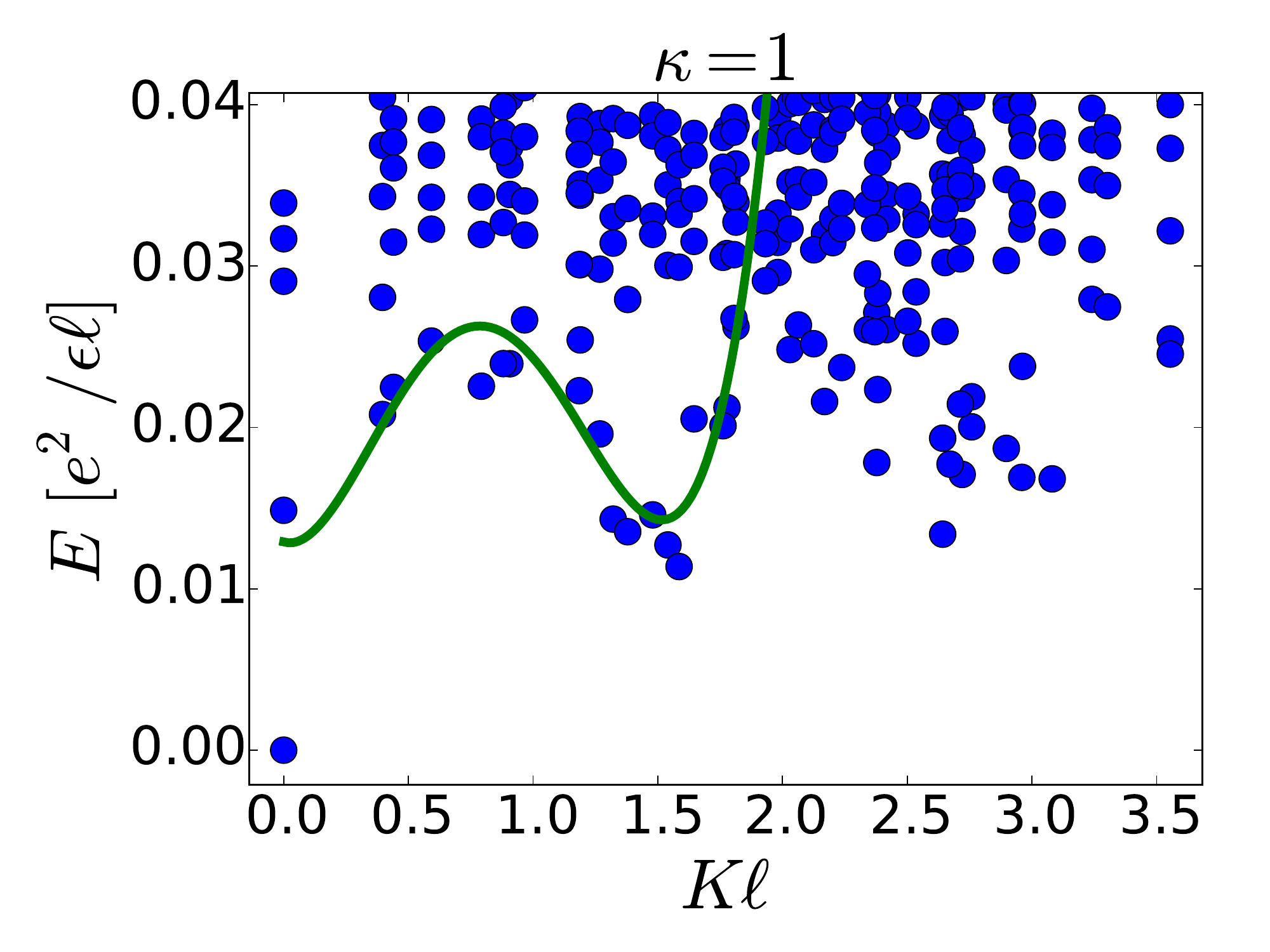} 
  \caption{Torus spectrum for $N_e=12$ electrons in the 2nd LL. Aspect ratio is 
0.9. Finite-width effects are modeled by a square well $w_{SQ}=4\ell$.
Here we display the full Brillouin zone for LL mixing parameter $\kappa=1$,
the largest value of Landau level mixing in perturbation theory.}
  \label{LL1}
\end{figure}

For comparison with experiments at $\nu=7/3$ it is 
important to know how and if the MR dispersion relation is changed by the 
inclusion of Landau level mixing
i.e. by virtual transitions of electrons towards the occupied LLL and 
unoccupied $N>1$ LLs.
This is an effect which is stronger in the second LL than in the LLL 
since the magnetic field is weaker and LL spacing is smaller and hence 
worth studying.
The strength of these effects is characterized by the ratio of the typical 
interaction energy and the cyclotron energy 
$\kappa=(e^2/\epsilon\ell)/(\hbar\omega_c)$ where $\omega_c=eB/m$ is
the cyclotron frequency. 
Bishara and Nayak~\cite{Bishara} have shown that integrating out virtual 
transitions to the LLL and higher levels with $N>1$ can be done in perturbation 
theory in the parameter $\kappa$.
At first order in $\kappa$ this procedure generates
additional two-body and three-body interactions~:
\be
\mathcal{H}_{eff}=\mathcal{H}_{Coulomb}+
\kappa\sum_{i<j}\sum_m\delta V^{(2)}_m \mathcal{P}^{(m)}_{ij}
+
\kappa\sum_{i<j<k}\sum_m\delta V^{(3)}_m \mathcal{P}^{(m)}_{ijk},
\ee
where $\mathcal{H}_{Coulomb}$ is the Coulomb Hamiltonian including finite-width 
effects projected onto the N=1 LL and the two-body and 
three-body pseudopotentials $\delta V^{(2)}_m,\delta V^{(3)}_m$
have been computed with a square well wavefunction in the $z$ 
direction~\cite{Peterson,Sodemann,Simon}, the operators 
$\mathcal{P}^{(m)}_{ij}$,
(resp. $\mathcal{P}^{(m)}_{ijk}$) are projectors onto the state of relative 
angular
momentum $m$ for two-body (resp. three-body) states.
These operators can be translated on the torus geometry and treated by exact 
diagonalization following ref.(\onlinecite{Pakrouski}).
It should be noted that the three-body interactions generate
an Hamiltonian matrix which is now much less sparse and there is thus a huge 
computational overhead. We have repeated the diagonalization of the largest
system with 12 electrons for $\kappa=0.5$ and $\kappa =1$
computing 5 eigenvalues in each momentum sector instead of 10. The 
lowest-lying 
energies are shown in Figs.(\ref{LL05},\ref{LL1}).
We find that while all 
energies are shifted by a 
$\kappa$-dependent value the MR dispersion relation remains essentially 
unaltered.
Notably
the characteristic wavevectors and gaps undergo only small changes for 
$\kappa\lesssim 1$, a value beyond which it is not clear one can trust 
perturbation theory.

So the main findings of the present work are robust to LL mixing at least 
within the scope of perturbation theory.

\section{Conclusions}
By torus exact diagonalization of the Coulomb interaction in 
the second 
Landau level for filling factor $\nu=7/3$ we have obtained the dispersion 
relation of the collective mode - the magnetoroton - expected from a liquid 
state
with Laughlin-like correlations. The observation of this mode requires both 
very large systems with more than 10 electrons and also a finite width of the 
electron gas. Even if this width reduces the gap scale as is the case in the 
LLL we find that the MR mode becomes discernible only when 
$w/\ell\gtrsim 1$. The Laughlin-like physics at $\nu=7/3$ is also seen in the 
guiding center structure factor which is almost isotropic.
The mode at long wavelength stays definitely below a higher-lying continuum of 
states. The gap from the ground state to the continuum is almost exactly twice 
the MR gap, probably indicating the two-particle nature of these states. While 
the MR 
minimum is at the same wavevector as in the LLL case there is also a maximum
of the dispersion at $k\ell\approx 0.8$. The MR mode disappears in the 
continuum for $k\ell\approx 1.8$. These features are captured correctly by the 
SMA and are resilient to Landau level mixing.

Several experiments have probed the MR in 
the LLL by inelastic light 
scattering~\cite{Pinczuk93,Pinczuk98,PinczukReview,Davies97,Kang2001,Kukushkin}
as well as phonon absorption~\cite{Mellor95,Zeitler99}
Detailed studies of the MR dispersion relation at $\nu=1/3$ are in quantitative 
agreement with theoretical calculations~\cite{Kang2001,Rhone2011} contrary to 
the magnetotransport 
gaps. Some details of the excited states beyond the lowest-lying MR branch are 
seen experimentally~\cite{Rhone2011,Hirjibehedin2005}. Recent works have 
started the study of the second Landau 
level physics where several phases are in competition beyond the FQHE 
liquids~\cite{Kumar2010,Kleinbaum2015,Wurstbauer2015,Levy2016,Jeong2016}.
The collective mode shape we observe from exact diagonalization and SMA 
calculations should be accessible to inelastic light scattering provided one 
uses a wide enough quantum well. Strictly speaking we cannot exclude that this 
shape is also correct for small width where finite-size effects are more 
problematic.
The existence of multiple critical points with vanishing derivatives in the 
dispersion should appear as several points with enhanced density of states.
\begin{acknowledgments}
We acknowledge discussions with M. Shayegan. We thank  
 IDRIS-CNRS Project 100383 for providing computer time allocations.
\end{acknowledgments}


\end{document}